\documentclass[aps,prl,twocolumn,superscriptaddress]{revtex4-1}
\usepackage{graphicx}
\usepackage{epstopdf}
\usepackage{color}
\usepackage{bm}
\usepackage{amsmath}
\usepackage{float}
\begin{document}

\title{Hydrogenic states of monopoles in diluted quantum spin ice}

\author{Olga Petrova}
\affiliation{Max Planck Institute for the Physics of Complex Systems, 01187 Dresden, Germany}
\author{Roderich Moessner}
\affiliation{Max Planck Institute for the Physics of Complex Systems, 01187 Dresden, Germany}
\author{S. L. Sondhi}
\affiliation{Max Planck Institute for the Physics of Complex Systems, 01187 Dresden, Germany}
\affiliation{Department of Physics, Princeton University, Princeton, NJ 08544, USA}

\begin{abstract}
We consider the effect of adding quantum dynamics to a
classical topological spin liquid, with particular view to how best to
detect its presence in experiment. For the Coulomb phase
of spin ice, we find quantum    effects to be most visible in the
gauge-charged monopole excitations. In the presence of weak dilution
with nonmagnetic ions we find a particularly crisp phenomenon, namely
the emergence of hydrogenic excited states in which a magnetic
monopole is bound to a vacancy at various distances. Via a mapping to
an analytically tractable single particle problem on the Bethe
lattice, we obtain an approximate expression for the dynamic neutron
scattering structure factor.
\end{abstract}

\maketitle

The quest for spin liquids is an important enterprise in strongly correlated many body physics in an era when 
a huge amount of theoretical interest has focused on forms of order outside the canonical broken symmetry paradigm \cite{Wegner,doi:10.1080/14786439808206568,WenNiu,doi:10.1080/00018739500101566}. The search involves identifying relatively simple Hamiltonians that host spin liquids and finding experimental
systems and signatures---the latter being more elusive than in Landau ordered systems. Indeed, at this point the list
of experimental systems where there is strong evidence of spin liquid behavior is small. Among them is the celebrated
spin ice system, arising in some rare earth pyrochlore magnets, which exhibits a U(1) spin liquid and excitations which
are condensed matter analogs of Coulombically interacting magnetic monopoles \cite{Castelnovo}. Spin ice is truly special 
at this point in hosting a three dimensional  spin liquid, but, owing to large magnetic moments, is limited to the classical regime,  in which coherent quantum dynamics appears to play little role. 

Logically, much recent interest has focused on looking for quantum generalizations of spin 
ice. There are several candidate materials for {\em quantum} spin ice behavior, such as ${\mathrm{Tb}}_{2}{\mathrm{Ti}}_{2}{\mathrm{O}}_{7}$ \cite{PhysRevLett.98.157204,PhysRevLett.99.237202,PhysRevLett.105.077203,PhysRevLett.109.017201}, ${\mathrm{Yb}}_{2}{\mathrm{Ti}}_{2}{\mathrm{O}}_{7}$ \cite{PhysRevLett.106.187202,PhysRevX.1.021002,NatCommun54970} and ${\mathrm{Pr}}_{2}{\mathrm{Zr}}_{2}{\mathrm{O}}_{7}$ \cite{Nakatsuji}, but an unambiguous experimental signature of quantum spin ice has been lacking.
Logically, much recent theoretical work has focused on looking for quantum generalizations of spin 
ice in which quantum fluctuations can lead to a fully quantum U(1) spin liquid  \cite{PhysRevB.68.184512,PhysRevB.69.064404,Henley, PhysRevB.86.104412, PhysRevLett.108.247210, PhysRevB.86.075154, arXiv:1410.0451} . 

Here we investigate the addition 
of quantum fluctuations to spin ice but in a different limit which is, plausibly, of relevance to existing materials. Fundamentally, we wish to understand the leading order effects of adding quantum dynamics about the classical spin ice limit. As we will detail below, this has a parametrically larger effect on monopole motion than on monopole-free ground states so the
leading manifestations of quantum fluctuations appear when monopoles are present.

We begin this program by studying the simplest manifestation of the quantum mechanics of monopoles---a striking effect that appears in the response of quantum spin ice to the 
introduction of a vacancy or missing spin. We find that the lowest lying excited states in the vicinity of the vacancy resemble those of hydrogen modulo lattice induced mixing---they involve a magnetic monopole bound to the impurity site into an infinite set of levels. In the presence of a dilute set of such impurities, these states give rise to a characteristic signature in neutron
scattering at low temperatures which we discuss. Readers may note the family resemblance of these hydrogenic monopole
states to hydrogenic states in doped semiconductors \cite[and references therein]{hydrogenicSC}, although we caution that the details have crucial differences. We also note that the response of spin liquids to impurities is of broad interest as a diagnostic of their internal dynamics: what happens  when you dope a spin liquid is the fundamental -- and to date largely unresolved -- question of the RVB theory of high temperature superconductivity \cite{Anderson1987}.

In the balance of the paper we begin by briefly reviewing how the dynamics of quantum spin ice can be formulated as the quantum mechanics of monopoles. We then concentrate our attention on the problem of a vacancy spin and describe
how it can be mapped to a good approximation to a monopole on a Bethe lattice interacting with a fixed Coulombic charge. This model leads to a family of hydrogenic bound states of the monopole along with a continuum band. 
In the technical heart of the paper we solve this problem and obtain an exact closed form solution for the onsite Green's functions. We use these results to obtain the signature of the hydrogenic states in the
structure factor of spin ice containing a dilute set of vacancy spins. We conclude with some comments and pointers to future
work. 

\noindent
{\bf Quantum dipolar spin ice:} Our model Hamiltonian 
\begin{equation}
H_\mathrm{QDSI}=H_\mathrm{DSI}+\sum_i {\bf t}\cdot{\bf S}_i \ 
\label{eq:QSIspin}
\end{equation}
consists, firstly, of the classical dipolar spin ice Hamiltonian, defined for Ising spins $ {\bf S}_i$ living on the sites of pyrochlore 
lattice and pointing along the local easy axis joining centers of neighboring tetrahedra \cite[and references therein]{Bramwell16112001,spinicereview}:
\begin{equation}
H_\mathrm{DSI}=\frac{\mu_0 \mu^2}{4 \pi} \sum_{i < j}\left[\frac{{\bf S}_i \cdot {\bf S}_j}{ r_{ij}^3}-\frac{3({\bf S}_i \cdot {\bf r}_{ij})({\bf S}_j \cdot {\bf r}_{ij})}{r_{ij}^5}\right].
\label{eq:QSIspinDip}
\end{equation}
The second term in Eq.~(\ref{eq:QSIspin}) is the transverse field, oriented perpendicular to the local easy axis, which adds the simplest quantum dynamics in the form of single spin flips. This simple form is convenient for a first theoretical analysis, for a more complete symmetry-based analysis of quantum terms in the Hamiltonian, see \cite{PhysRevB.78.094418,PhysRevB.86.104412}.

\noindent
{\bf Ghost spins and the Bethe lattice:} At this point we switch from a spin description to that referred to as the \emph{dumbbell model} \cite{Castelnovo}
which we quickly review. Each spin is replaced by a pair of magnetic charges $\pm q_m=\mu/a_d$ of opposite sign, where $\vec{a}_d$ is a vector pointing between the centres of neighboring tetrahedra.
By summing up the net charge at the center of each tetrahedron ($Q_\alpha \equiv \sum_{i \in \alpha} q_i = 0, \pm 2 q_m, \pm 4 q_m$), we can replace the dipolar piece of the
spin Hamiltonian (\ref{eq:QSIspinDip}) by 
\begin{equation}
H =
\frac{\mu_0}{4 \pi} \sum_{\alpha < \beta} \frac{Q_\alpha Q_\beta}{r_{\alpha\beta}}
+
\frac{v_0}{2} \sum_\alpha Q_\alpha^2 \ ,
\label{eq:dumbH}
\end{equation}
Coulomb interactions between charges on the diamond lattice, 
with $v_0=\frac{\mu_0}{4\pi}\frac{2}{a}\left( 1 + \sqrt{\frac{2}{3}} \right)$ the cost of creating a monopole, 
which can be shifted by  a nearest-neighbor exchange term. In spin ice ground states, $Q_\alpha\equiv0$. 
Flipping a spin in a ground states yields a pair of magnetic monopoles of charges $\pm2q_m$  on
adjoining tetrahedra which can then move apart via further spin flips at finite cost in energy. Each charge has three \emph{majority} spins that are all pointing in or out of the tetrahedron, and a single \emph{minority} spin pointing in the opposite direction.

\begin{figure}		
\includegraphics[width=0.83\linewidth]{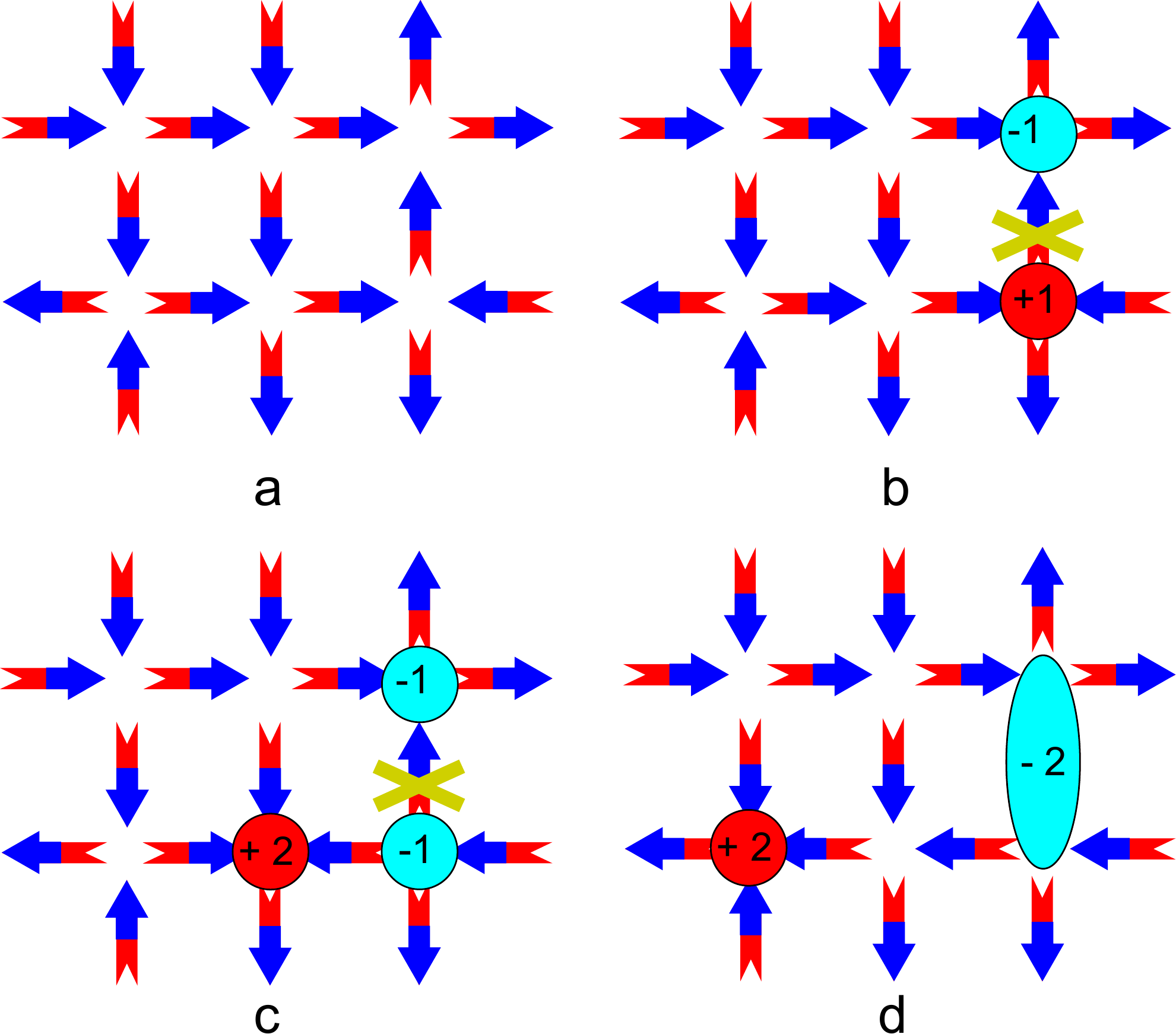}
\caption{Spin ice projected onto a plane, with each vertex of the resulting square lattice in (a) corresponding to a pyrochlore tetrahedron. (b) A missing spin gives rise to a $+1$ and a $-1$ charges; (c) flipping one of the majority spins adjacent to the vacancy creates a bulk charge $+2$, inverting the sign of one of the vacancy charges; (d) the bulk charge propagates 
in the system through further spin flips, while the net charge of vacancy is approximated as a single $-2$ monopole.}
\label{fig:vacancy}
\end{figure}

With this in hand let us discuss the energetics of substituting a magnetic ion on the pyrochlore lattice by a non-magnetic impurity,
see Fig.~\ref{fig:vacancy}. Removing a spin from a classical spin ice state (Fig.~\ref{fig:vacancy}a) leaves behind two 
monopoles of charges $\pm q_m$  \cite{arXiv:1405.0668} (Fig.~\ref{fig:vacancy}b). A bulk monopole with 
charge $\pm2q_m$ can be `emitted' by the vacancy via flipping one of the two majority spins at each of the tetrahedra adjacent to the ghost spin (Fig.~\ref{fig:vacancy}c). The bulk monopole is then free to move around in the system. We define a quantity that can be thought of as the ionization energy of the vacancy: $I=\frac{\mu_0}{2\pi}\frac{\mu^2}{a_d^2}+2v_0\frac{\mu^2}{a_d^2}$. This is the total energy cost of  an emitted monopole moved out to infinity. Emitting another monopole into the bulk would cost additional energy of the order of $2v_0\frac{\mu^2}{a_d^2}$, so the lowest energy charged excitation in the presence of an isolated vacancy is a single monopole with charge $\pm2q_m$. Once the monopole is emitted into the bulk, it is free to hop through fluctuation-induced spin flips,  while a net charge of opposite sign $\mp2q_m$ remains at the vacancy (Fig.~\ref{fig:vacancy}d). 

Adding quantum dynamics via the transverse field in Eq.~(\ref{eq:QSIspin}) has only a weak effect on the ground states,
as  connecting two of them requires flipping spins in closed loops, {\it minimally} 
six of them on a hexagon of the pyrochlore lattice. Near the classical limit,  $v_0\gg t$,
such processes come with a prohibitively small energy scale, $\sim t^6/v_0^5$. 
By contrast, for a state containing a monopole, the lowest order effect -- a monopole hopping onto a neighboring 
tetrahedron by flipping a majority spin -- is parametrically stronger: {\it  
linear} in $t$!
 
Thus, in experiment, the most promising place to see quantum effects in spin ice is in the gauge-charged monopole 
excitations, rather than its gauge-neutral gapless emergent photons. Analogous
considerations apply in the proximity of a vacancy, where we focus on the case of a monopole  emitted into the bulk (Fig.~\ref{fig:vacancy}d), also with low-order signatures. 
For this reason, here we perform a quantum calculation for the monopole states, and do a thermal sum over the nearly degenerate spin ice configurations. 

We treat the problem as that of two Coulombic charges, one of which is stationary. As the charge propagates through the bulk, it changes the spin ice background. This process is difficult to capture exactly, but fortunately it is possible to make considerable progress via an effective model that we describe next. From this, we are able to extract the 
bound states in considerable detail,  followed by a continuum band, much as we would expect for the Hydrogen atom. 

\begin{figure}	
\vspace{5mm}	
\includegraphics[width=0.75\linewidth]{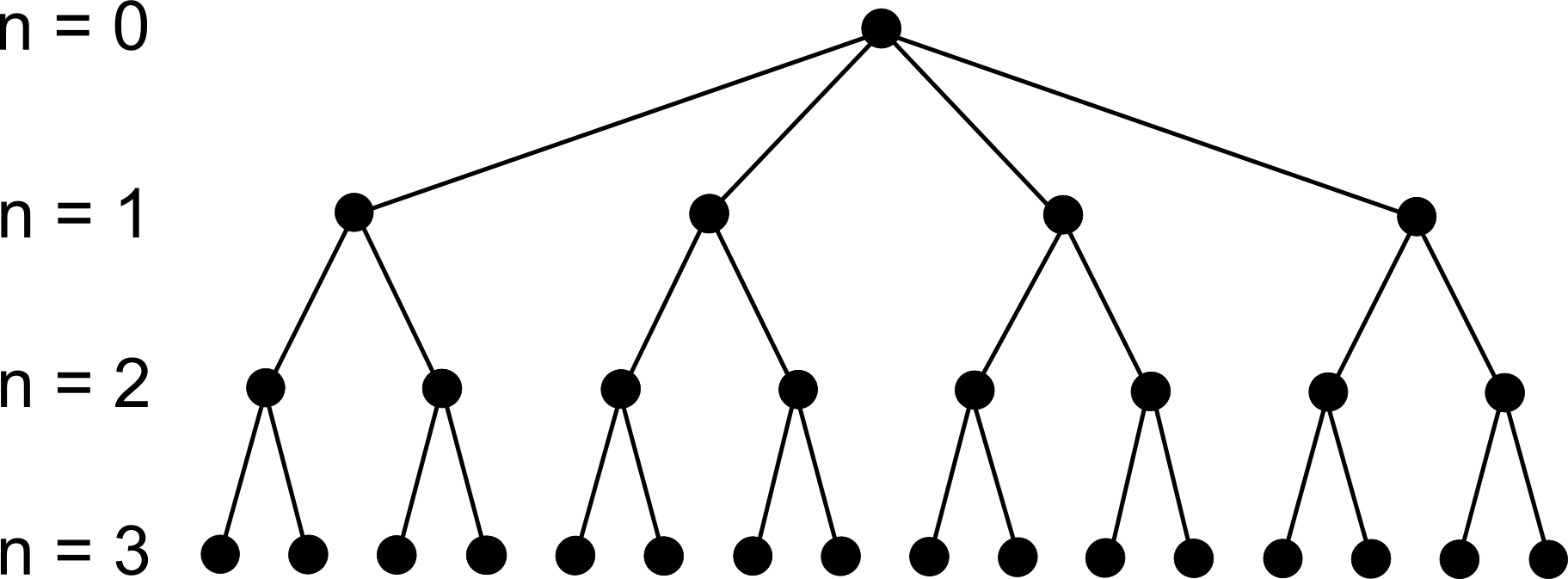}
\caption{The Bethe lattice describing state space. Its root,  $n=0$, corresponds to the unionized vacancy state $|0\rangle$.}
\label{fig:bethe}
\end{figure}

In order to investigate the problem of an isolated vacancy that has emitted a free monopole into the bulk, we switch to the \emph{state lattice} description \cite{Chen}. First, consider a new basis of the following (classical) states: a spin ice state with a vacancy, which we label $|0\rangle$, and states with an emitted monopole in the bulk, connected to $|0\rangle$ through single spin flips. Next, each site of the state lattice represents one of the basis states $|n\rangle$; while bonds connect those sites 
whose corresponding states are connected by single spin flips. Apart from site 0 (representing $|0\rangle$), the state lattice is trivalent. It can be shown that the smallest closed cycle in the state lattice of disordered pyrochlore spin ice has length 20 \cite{Supplement}. We therefore approximate the state lattice by a cycle-free infinite Cayley tree (the Bethe lattice) rooted at site 0
(Fig.~\ref{fig:bethe}). The monopole propagating in real space 
corresponds to
a single particle hopping on this lattice in the presence of the Coulomb potential:
\begin{equation}
H|0\rangle  =-t\sum_{m=1}^4|m\rangle;\quad H|n\rangle  =\left(I+\frac{C}{d_n}\right)|n\rangle-t\sum_{m=1}^3|m\rangle\quad 
\label{eq:QSIstates}
\end{equation}
%
where the sums run over states reached by flipping majority spins of the monopole.
$d_n$ denotes the distance between the vacancy and the monopole in the bulk in units of $a_d$, such that $C/d_n$ is the attractive Coulomb potential ($C=-\frac{\mu_0}{\pi}\frac{\mu^2}{a_d^2}$) between the two charges. In conventional spin ice, the cost of having a monopole is larger than the magnitude of the Coulomb interaction between two charges,  $I>|C|$. For concreteness, we use $C=-I/3$; $t=I/10$ in the following. Since we restrict ourselves to a particular starting spin ice configuration and omit other degenerate ice states from the discussion, the mapping from Eq.~(\ref{eq:QSIspin}) to (\ref{eq:QSIstates}) is accurate to $\mathcal{O}(t^5/v_0^4)$. Our final approximation concerns the distance between monopoles. Since the four sites at the first generation of the Bethe lattice correspond to the bulk monopole being one spin flip away from the vacancy, it is natural to approximate $d_n$ in  Eq.~(\ref{eq:QSIstates}) by the generation of the Bethe lattice $n$. 
This definition fails to be exact already beyond $\mathcal{O}(t^2)$, but should work sufficiently well in the $I,|C|\gg t$ regime, when the bulk monopole prefers not to move too far. In return for these approximations, we are able to solve exactly our idealized model, that of a single particle hopping on the Bethe lattice in the presence of a Coulomb potential $I+C/n$ for $n>0$.

\begin{figure}		
\includegraphics[width=\linewidth]{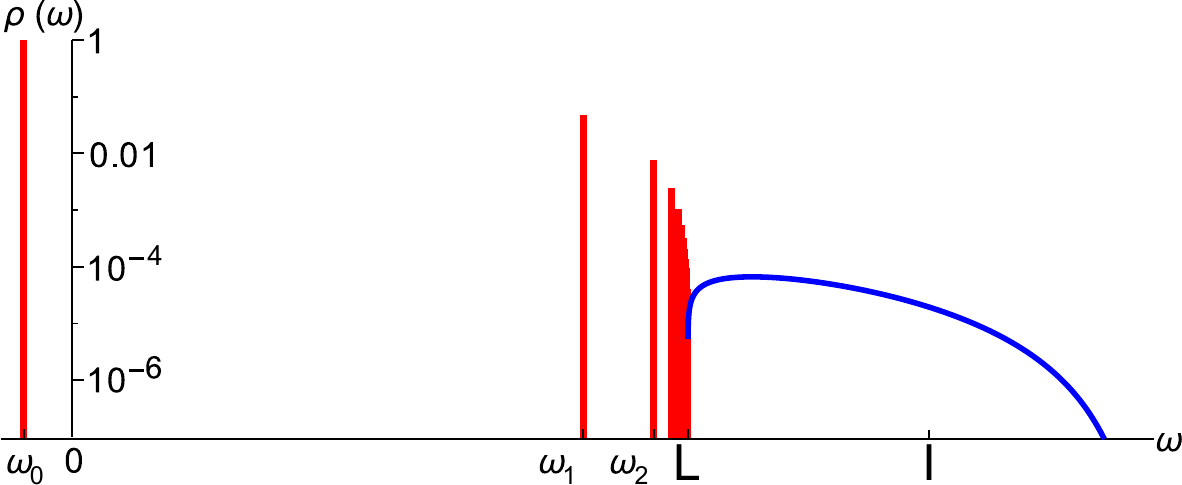}
\caption{Local density of states at site 0 of the Coulomb problem on the Bethe lattice, for $C=-I/3$; $t=I/10$. Bound states (red) appear as sharp peaks; the lower edge of the continuum (blue) is labeled $L$, the classical ground state energy by $0$ and the  ionization energy by $I$. }
\label{fig:localDOS}
\end{figure}

\noindent
{\bf The Bethe lattice problem:} We calculate the diagonal elements $\mathcal{G}_{ii}(\omega)$ of the lattice Green's function to infinite order in $t$ \cite{Brinkman,Gallinar}. We find \cite{Supplement} that each $\mathcal{G}_{ii}(\omega)$ can be written down in terms of a finite number of $\mathcal{G}_{k}^{F}(\omega)$, infinite sums involving particle hopping from a site at generation $k$ to sites at generations $g>k$. The latter have a closed form expression in terms of the Gauss hypergeometric functions $F^2_1(a,b,c,z)$ \cite{Ramanujan}:
\begin{widetext}
\begin{equation}
\mathcal{G}^{F}_k(\omega)=\frac{2k/\omega}{\sqrt{1+x^2}+1}
\frac{1}{k-\frac{C/\omega}{\sqrt{1+x^2}}}
\frac{F^2_1\left(1-\frac{C/\omega}{\sqrt{1+x^2}},k+1,k+1-\frac{C/\omega}{\sqrt{1+x^2}},\frac{1-\sqrt{1+x^2}}{1+\sqrt{1+x^2}}\right)
}{F^2_1\left(1-\frac{C/\omega}{\sqrt{1+x^2}},k,k-\frac{C/\omega}{\sqrt{1+x^2}},\frac{1-\sqrt{1+x^2}}{1+\sqrt{1+x^2}}\right)}
\label{eq:Gexact}
\end{equation}
\end{widetext}
where $x^2=-\frac{8t^2}{\omega^2}$.  This yields the exact expression for any of the diagonal elements of the Green's function; for instance at the root site
\[
\mathcal{G}_{00}(\omega)=\left(\omega-4t^2\mathcal{G}^{F}_1(\omega-I)\right)^{-1}.
\]
The full Green's function yields the energy levels via its poles and the local densities of states for each Bethe lattice generation, proportional to its imaginary part. The local density of states at site 0 in Fig.~\ref{fig:localDOS} indicates that indeed there are bound states followed by the continuum energy band. While the classical ground state (a spin ice state with a vacancy) would have zero energy, the ground state energy of the quantum problem $\omega_0$ is lowered due to the hopping $t$. Low-lying excited states are separated from the ground state by a gap, which is also decreased from the classical value $I$ through hopping and Coulomb attraction. They accumulate below the edge of the continuum band, located at $L=I-\sqrt{8t^2}$. In the Bethe lattice problem, the band of the extended states, of width linear in $t$, is confined to the region $I-\sqrt{8t^2}<\omega<I+\sqrt{8t^2}$. (Introducing closed cycles into the lattice has the effect of adding band tails, extending beyond these edges.)

\begin{figure}
\includegraphics[width=0.9\linewidth]{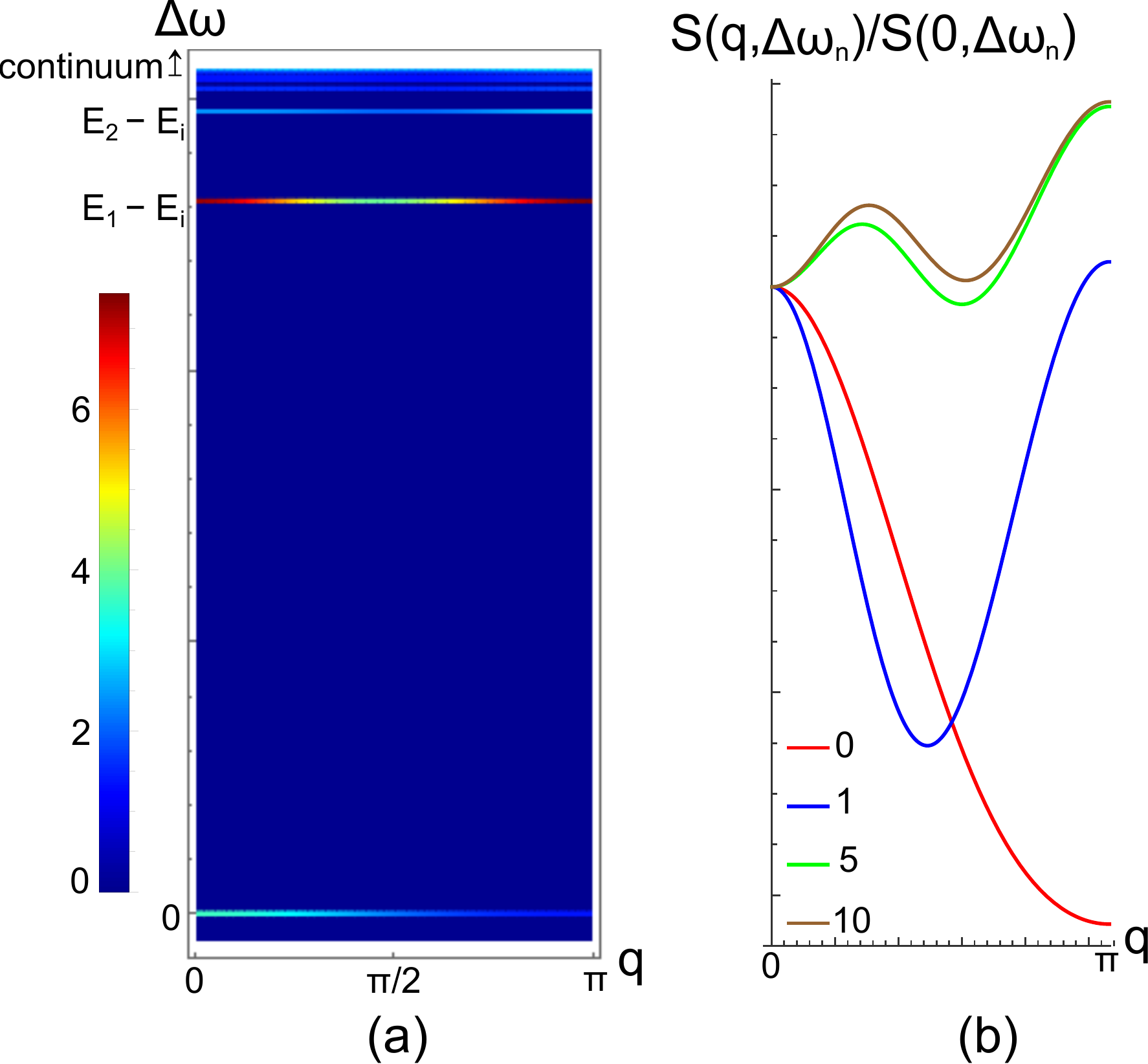}
\caption{Left: dynamic structure factor $S(q,\Delta\omega)$ for powder averaged neutron scattering. Each line is multiplied by Exp$\left[0.7|n-1|\right]$, where $n$ labels the all-even energy levels  with $n=0$  the ground state. Right:  Lineshapes $S(q,\Delta\omega_n)/S(0,\Delta\omega_n)$ for $n=0,1,5,10$. $C=-I/3; t=I/10$ throughout.}
\label{fig:Sqw}
\end{figure}

\noindent
{\bf Signatures of monopoles in neutron scattering:} One of our central results is the dynamic structure factor, defined as
\begin{equation}
S(\vec{q},\Delta\omega)=\sum_f \delta(E_f-E_i-\Delta\omega)\lvert{\sum_{\vec{R}}\langle f\lvert S^{+}_{\vec{R}}\rvert i\rangle e^{i\vec{q}\cdot\vec{R}}}\rvert^2.
\label{eq:Scattering}
\end{equation}
In order to extract the information that is most relevant to spin ice experiments from the Bethe lattice model, we calculate a one dimensional version of Eq.~(\ref{eq:Scattering}), averaged over all directions of $\vec{q}$. Such a quantity, $S(q,\Delta\omega)$, can be measured directly in a powder averaged neutron scattering experiment.  The details of our calculation, carried out in the limit of dilute nonmagnetic impurities, are given in the Supplemental Material \cite{Supplement}. The dynamic structure factor $S(q,\Delta\omega)$, plotted in Fig.~\ref{fig:Sqw}(a), has sharp features signaling the presence of bound states. The structure of the lines gives direct information about the character of the ground and excited states. The most visible signatures show up in elastic scattering and at the energy transfer equal to the difference between the first excited state and the ground state. For a well-localized ground state, the matrix elements between $S^{+}_{\vec{R}}\rvert i\rangle$ and excited states at higher energies (bound to the vacancy at distant radii) give rise to peaks whose structure is essentially identical up to a scale factor, as shown in Fig.~\ref{fig:Sqw}(b). Note that for $t=0$, the signals corresponding to $n\neq1$ would be absent, vanishing as powers of $t$. Their presence thus yields direct evidence of the existence of quantum dynamics.

\noindent
{\bf Conclusion and outlook:} We have studied in detail the properties of 
magnetic monopoles in dipolar quantum spin ice. We have demonstrated that these are the prime 
indicators of the presence of quantum dynamics. In
the presence of nonmagnetic impurities we have found both sharp 
hydrogenic bound states as well as  a broad continuum energy band. While we believe these results to be robust,
there is clearly much scope for further, presumably numerical, modeling
taking into account the detailed lattice structure, as well as any
material specific single-ion physics and terms in the quantum Hamiltonian.

The quantum dynamics of a  pair of monopoles presents a more difficult problem  due to the pair's center of mass motion. We are planning to address this issue, as well as clarify the detailed character of the continuum band of states in the vacancy problem, in future work. Additionally, despite neutron scattering being the method of choice for investigating magnetic materials, local disorder is an attractive subject for other types of experimental probes, such as nuclear magnetic resonance. While such techniques are beyond the scope of this work, our theoretical model  can also be employed for calculating real space quantities accessible by the local probes.

\emph{Acknowledgments:} The authors thank Subhro Bhattacharjee, Claudio Castelnovo, Radu Coldea, Siddhardh Morampudi, Satoru Nakatsuji, and Oleg Tchernyshyov for useful discussions, and Yen Ting Lin and Nikos Bagis for pointing out the relevance of Ramanujan's work to the Green's function calculation. The authors acknowledge support of the Helmholtz Virtual Institute \emph{New States of Matter and their Excitations}, the US National Science Foundation under Grant No. DMR-1311781 (SS), the Alexander von Humboldt Foundation (OP and SS), and the German Science Foundation (DFG) via the Gottfried Wilhelm Leibniz Prize Programme.

\bibliography{spinicebib}

\begin{widetext}

\section*{Supplementary Material}

\subsection*{Minimal closed cycle in the state graph of pyrochlore spin ice}

\begin{figure}[H]
\begin{center}
\includegraphics[width=0.25\linewidth]{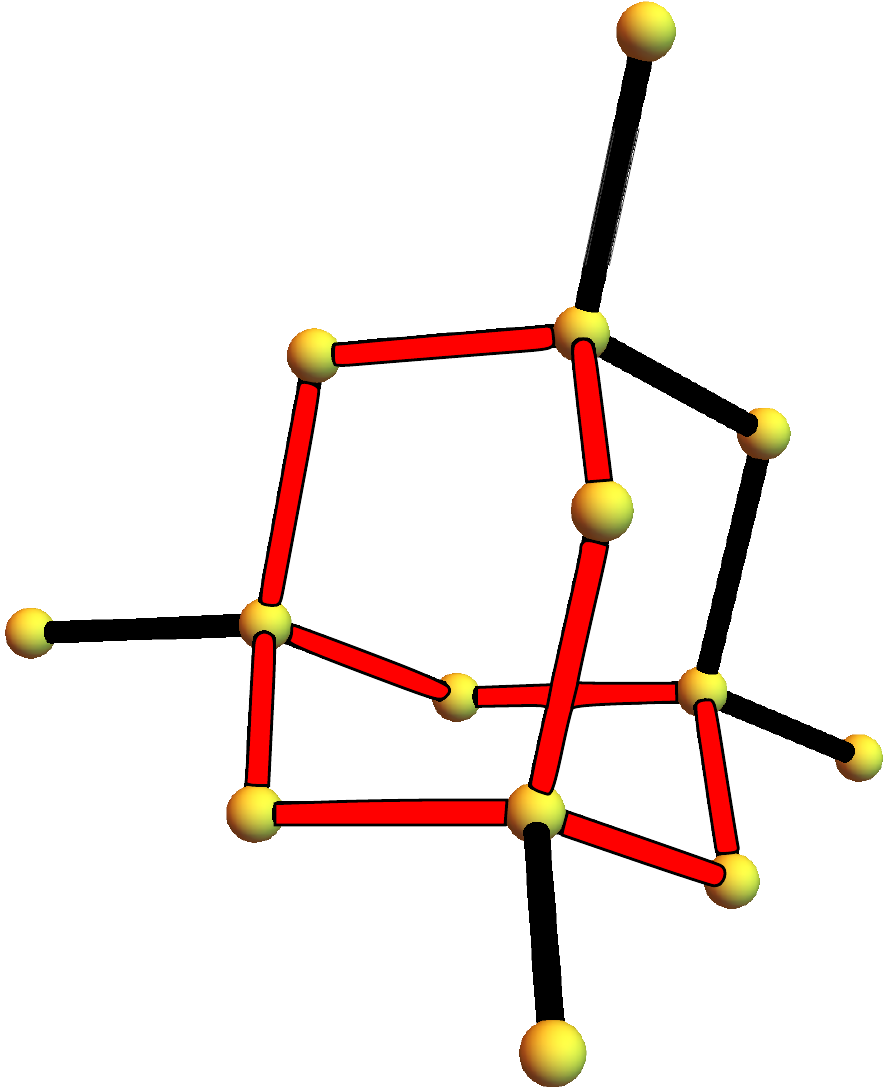}
\includegraphics[width=0.25\linewidth]{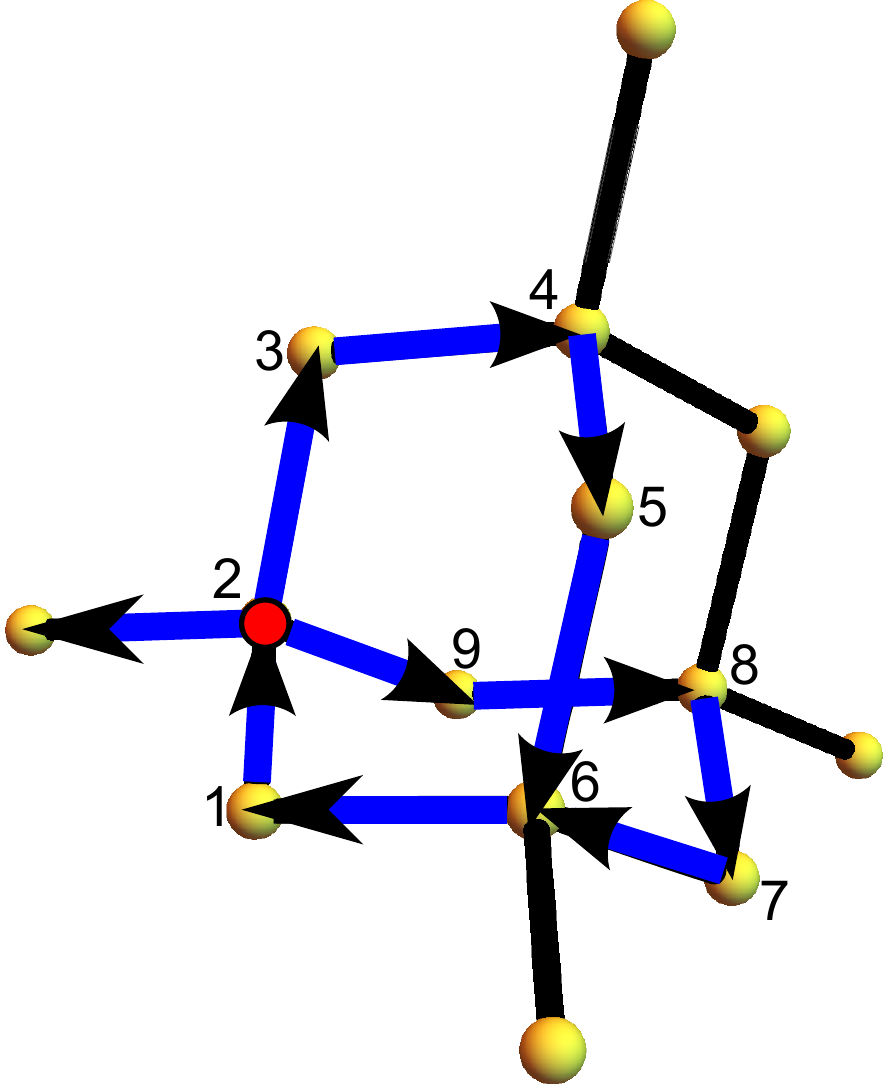}
\end{center}
\caption{A fragment of the real-space diamond lattice that the magnetic monopoles of pyrochlore spin ice reside on. Spins are depicted as arrows pointing along the lattice edges. Left: the shortest closed cycle in the state graph involves flipping spins along two hexagons that share two edges in common (red). Right: an example of a spin configuration, that gives rise to the shortest possible cycle on the state lattice, with two flippable hexagons and a magnetic monopole at one of the sites where they intersect.}
\label{fig:closed}
\end{figure}

One of the arguments for using the Bethe lattice to approximate the state graph of pyrochlore spin ice is that the closed cycles present in the actual state lattice are long and thus can be neglected. The spins can be thought of as pointing along the edges of the diamond lattice \cite{Castelnovo}. Its smallest closed loop has length six, so inverting spins along a \emph{flippable} hexagon (such that six spins are arranged to point head-to-tail) brings the charge back to where it started. The spin configuration is changed by such a process, and one needs to flip the same six spins a second time to bring the system back to the initial state. However, this process constitutes a self-retracing path on the state graph, rather than a closed cycle.

Introducing a loop into the state graph requires adding another hexagon into the real-space picture. The shortest possible loop is obtained by maximizing the overlap between the two hexagons: they share two edges as shown in Fig.~\ref{fig:closed}. A non-trivial closed loop on the state graph involves ten spins on two such overlapping hexagons, and positioning a magnetic monopole on a site from which it can hop along two different paths [charge $-2q_m$ on site 2 in Fig.~\ref{fig:closed}]. Each spin can be labeled by two numbers that correspond to the diamond lattice sites at the two ends of the dumbbell. An example of a closed cycle in the state graph, written down as a sequence of spin flips as labeled in Fig.~\ref{fig:closed}, is then:

\[
\langle23\rangle\langle34\rangle\langle45\rangle\langle56\rangle\langle61\rangle\langle12\rangle
\langle29\rangle\langle98\rangle\langle87\rangle\langle76\rangle\langle65\rangle\langle54\rangle
\langle43\rangle\langle32\rangle\langle21\rangle\langle16\rangle\langle67\rangle\langle78\rangle
\langle89\rangle\langle92\rangle.
\]
This closed cycle has length 20. We would like to emphasize that this is the shortest possible loop in the state lattice of \emph{disordered} classical ice states. Other spin configurations have closed cycles that either have the same length, or are even longer.

\subsection*{Lattice Green's function: the continued fraction method}

In order to solve the Coulomb problem on the Bethe lattice, we employ the continued fraction approach \cite{Brinkman,Gallinar}. The strategy is to start with a perturbative treatment (where $H_0$ is the potential, and nearest neighbor hopping $t$ is the perturbation) and proceed to solve the problem exactly by calculating the self-energy to infinite order in $t$ in the particle's lattice Green's function. Recursively solving Dyson's equation
\[
\mathcal{G}(\omega)=\mathcal{G}_0(\omega)+\mathcal{G}_0(\omega)\Sigma(\omega)\mathcal{G}(\omega)
\]
(where $\Sigma(\omega)$ is the self energy) gives rise to continued fractions for the Green's function. For instance, the diagonal element of the Green's function for site 0 is:
\[
\mathcal{G}_0(\omega)=\frac{1}{\omega-\Sigma_0(\omega)}=\cfrac{1}{\omega-\cfrac{4t^2}{\omega-I-C-\cfrac{2t^2}{\omega-I-\frac{C}{2}-...}}}.
\]
The self energy $\Sigma_0(\omega)$ above is given by the sum of all paths on the lattice going away from site 0 and back to it. The elements of the Green's function for sites at all generations can be defined in terms of a finite number of $\mathcal{G}_{k}^{F}(\omega)$, infinite continued fractions involving hops only to generations higher than $k$:
\[
\mathcal{G}_{k}^{F}(\omega)=\cfrac{1}{\omega-\frac{C}{k}-\cfrac{2t^2}{\omega-\frac{C}{k+1}-\cfrac{2t^2}{\omega-\frac{C}{k+2}-...}}}.
\label{eq:Gcont}
\]
Using a theorem by Ramanujan \cite{Ramanujan}, we arrive at the following exact closed form expression for the continued fraction above:
\[
\mathcal{G}^{F}_k(\omega)=\frac{2k/\omega}{\sqrt{1+x^2}+1}
\frac{1}{k-\frac{C/\omega}{\sqrt{1+x^2}}}
\frac{F^2_1\left(1-\frac{C/\omega}{\sqrt{1+x^2}},k+1,k+1-\frac{C/\omega}{\sqrt{1+x^2}},\frac{1-\sqrt{1+x^2}}{1+\sqrt{1+x^2}}\right)
}{F^2_1\left(1-\frac{C/\omega}{\sqrt{1+x^2}},k,k-\frac{C/\omega}{\sqrt{1+x^2}},\frac{1-\sqrt{1+x^2}}{1+\sqrt{1+x^2}}\right)}
\]
where $x^2=-\frac{8t^2}{\omega^2}$ and $F^2_1(a,b,c,z)$ is the Gauss hypergeometric function. 

\subsection*{Calculation of the dynamic structure factor}

\subsubsection*{Quantum numbers of the Bethe lattice eigenstates}

Consider a set of only three sites: site 1 at generation $n$ and sites 2 and 3, connected to it, at generation $(n+1)$. Exchanging sites 2 and 3 leaves the Hamiltonian (\ref{eq:QSIstates}) invariant. Now examine what happens when Eq.~(\ref{eq:QSIstates}) acts on two states: $|\Psi_S\rangle$ and $|\Psi_A\rangle$, symmetric and antisymmetric combinations of the particle being at sites 2 and 3. The diagonal part of the Hamiltonian will be the same for both states, and both states allow for the particle to hop to higher generation sites. However, hopping back to site 1 is eliminated by $|\Psi_A\rangle$, since the contributions from sites 2 and 3 cancel out. Similarly, each Bethe lattice generation $n\ge 1$ can be assigned an even/odd quantum number, corresponding to the symmetries that involve exchanging left and right sites at each generation. If a particle starts at a state that is odd at $k^{th}$ generation, it can only hop to sites at generations $n>k$. It follows that the particle's wavefunction has zero amplitude at all generations $n<k$. Therefore, only the all even states have a nonzero amplitude at the origin.

\subsubsection*{Powder averaged dynamic structure factor}

The dynamic structure factor is defined as
\[
S(\vec{q},\Delta\omega)=\sum_f \delta(E_f-E_i-\Delta\omega)\lvert{\sum_{\vec{R}}\langle f\lvert S^{+}_{\vec{R}}\rvert i\rangle e^{i\vec{q}\cdot\vec{R}}}\rvert^2.
\]
We calculate the exact expression for $S(\vec{q},\Delta\omega)$ that is averaged over all directions of $\vec{q}$ for the Bethe lattice problem. This gives us the approximate dynamic structure factor $S(q,\Delta\omega)$ that can be measured in a powder averaged neutron scattering experiment on quantum spin ice, in the limit of weak dilution.

The ground state $\rvert i\rangle$ of the Coulomb problem on the Bethe lattice can be written down in the following form:
\[
\rvert i\rangle=\alpha\rvert 0\rangle+\beta\rvert 1\rangle+\gamma\rvert 2\rangle+\dots
\]
where the Greek letters represent the ground state's amplitude per lattice generation. In the course of a scattering experiment, a neutron coming in flips a spin. Since we restrict ourselves to having at most one free monopole with charge $\pm2q_m$, the spin flip either nucleates a monopole at the vacancy, eliminates it, or shifts an existing monopole by a lattice spacing:
\[
\sum_{\vec{R}}e^{i\vec{q}\cdot\vec{R}}S^{+}_{\vec{R}}\rvert i\rangle = \alpha \left( 4e^{iq}\rvert 1\rangle\right)+\beta\left(e^{iq}\rvert 0\rangle+2e^{i2q}\rvert 2\rangle\right)+\gamma\left(e^{i2q}\rvert 1\rangle+2e^{i3q}\rvert 3\rangle\right)+\dots
\]
Since the ground state belongs to the all even sector, the signs of the amplitudes for the sites at the same generation are equal, and the contributions from the spin flips add up (hence the numerical prefactors in the expression above). When we take the inner product of $\sum_{\vec{R}}e^{i\vec{q}\cdot\vec{R}}S^{+}_{\vec{R}}\rvert i\rangle$ with excited states which contain odd quantum numbers, however, the hops to sites at the same generation cancel out. Therefore, only the all even states contribute to the powder averaged structure factor. We can see how these states are distributed in real space by plotting their probabilities as a function of Bethe lattice generations $n$ (Fig.~\ref{fig:statesform}). In particular, the ground state is localized near the origin, which leads to the scattering intensity being high for lower energy states, and low for the states that accumulate near the continuum band edge, whose wavefunctions near the vacancy differ in amplitude, but not much in shape.

\begin{figure}		
\includegraphics[width=0.45\linewidth]{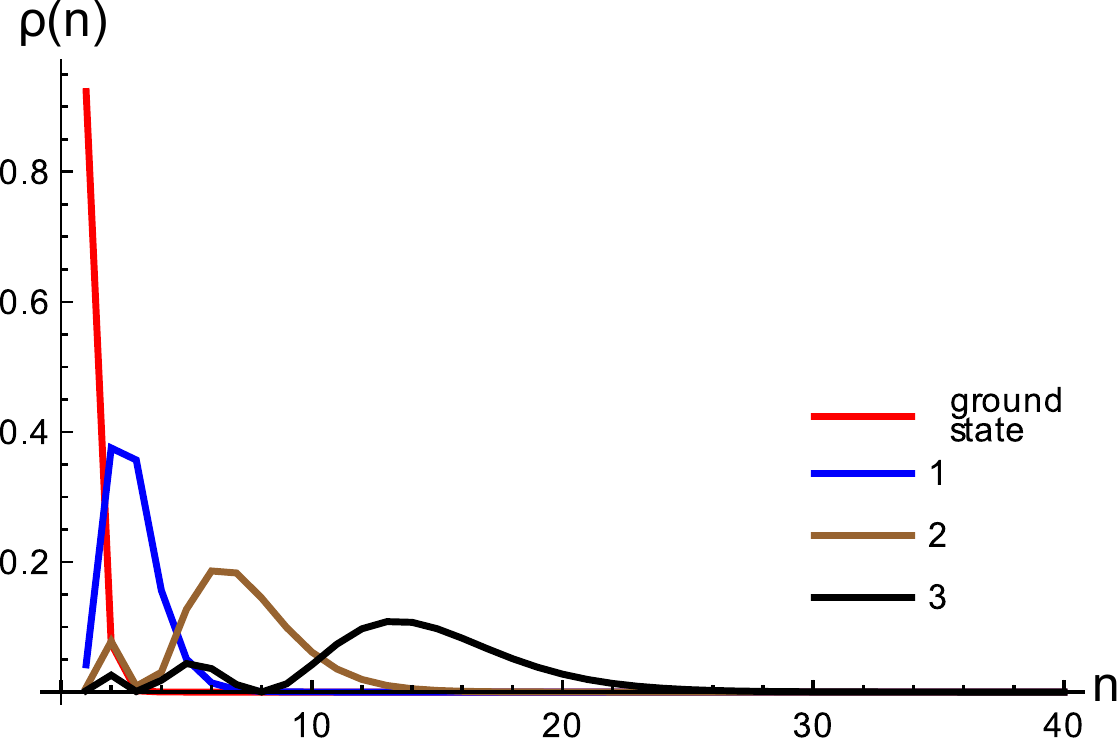}
\caption{Probabilities of the four lowest energy states in the all even sector, plotted as a function of the Bethe lattice generation. $C=-I/3; t=I/10$.}
\label{fig:statesform}
\end{figure}

\end{widetext}
\end{document}